\documentclass[aps,prl,twocolumn,showpacs]{revtex4}

\usepackage{bm,bbm,amsmath, amssymb,amsbsy,graphicx,amsthm,color,amsthm,hyperref,txfonts}

\newcommand{\kebra}[2]{\vert{#1}\rangle\langle{#2}\vert}
 \newcommand{\tr}[1]{\text{Tr}}
\newcommand{\ket}[1]{|#1\rangle}
\newcommand{\bra}[1]{\langle#1|}

\newcommand{\med}[1]{\langle{#1}\rangle}

\newtheorem*{Theorem}{Theorem}
\newtheorem{Lemma}{Lemma}
 \newtheorem*{Corollary}{Corollary}

\newtheorem*{Rem}{Proposition}
\begin{document}

\title{Characterizing Nonclassical Correlations via Local Quantum Uncertainty}

\author{Davide Girolami$^1$, Tommaso Tufarelli$^2$, and Gerardo Adesso$^1$}
\affiliation{$^1\hbox{School of Mathematical Sciences, The University of Nottingham, University Park, Nottingham NG7 2RD, United Kingdom}$ \\
$^2\hbox{QOLS, Blackett Laboratory, Imperial College London, London SW7 2BW, United Kingdom}$ }

\begin{abstract}
Quantum mechanics predicts that measurements of incompatible observables carry a minimum uncertainty which is independent of technical deficiencies of the measurement apparatus or incomplete knowledge of the state of the system. Nothing yet seems to prevent a single physical quantity, such as one spin component, from being measured with arbitrary precision. Here we show that an intrinsic quantum uncertainty on a single observable is ineludible in a number of physical situations. When revealed on local observables of a bipartite system, such uncertainty defines an entire class of {\it bona fide} measures of nonclassical correlations. For the case of $2\times d$ systems, we find that a unique measure is defined, which we evaluate in closed form. We then discuss the role that these correlations, which are of the `discord' type, can play in the context of quantum metrology. We show in particular that the amount of discord present in a bipartite mixed probe state guarantees a minimum precision, as quantified by the quantum Fisher information, in the optimal phase estimation protocol.
 \end{abstract}

\date{May 20, 2013}


\maketitle

\noindent {\it Introduction.}---
In a classical world, error bars are exclusively due to technological limitations, while quantum mechanics entails that two noncommuting observables cannot be jointly measured with arbitrary precision \cite{heis}, even if one could access a flawless measurement device. The corresponding uncertainty relations have been linked to distinctive quantum features such as nonlocality, entanglement and data processing inequalities \cite{opp,unc,coles}.

Remarkably, even a single quantum observable may display an intrinsic uncertainty as a result of the probabilistic character of quantum mechanics.   Let us consider for instance a composite system prepared in an entangled state \cite{entanglement}, say the Bell state $|\phi^{+}\rangle =\frac{1}{\sqrt 2}(|00\rangle +|11\rangle)$ of two qubits. This is an eigenstate of the global observable $\sigma_z\otimes\sigma_z$ ($\vec\sigma=(\sigma_x,\sigma_y,\sigma_z)$ are the Pauli matrices), so there is no uncertainty on the result of such a measurement.  On the other hand, the measurement of {\it local} spin observables of the form $\vec a\cdot\vec\sigma\otimes\mathbb I$  (where $\vec a \neq0$ is a real vector) is intrinsically uncertain. Indeed, the state $|\phi^{+}\rangle\bra{\phi^{+}}$, and in general any entangled state, cannot be eigenstates of a local observable. Only uncorrelated states of the two qubits, e.g.~$|00\rangle$, admit at least one completely `certain' local observable.

\begin{figure}[t!]
\includegraphics[width=5.5cm]{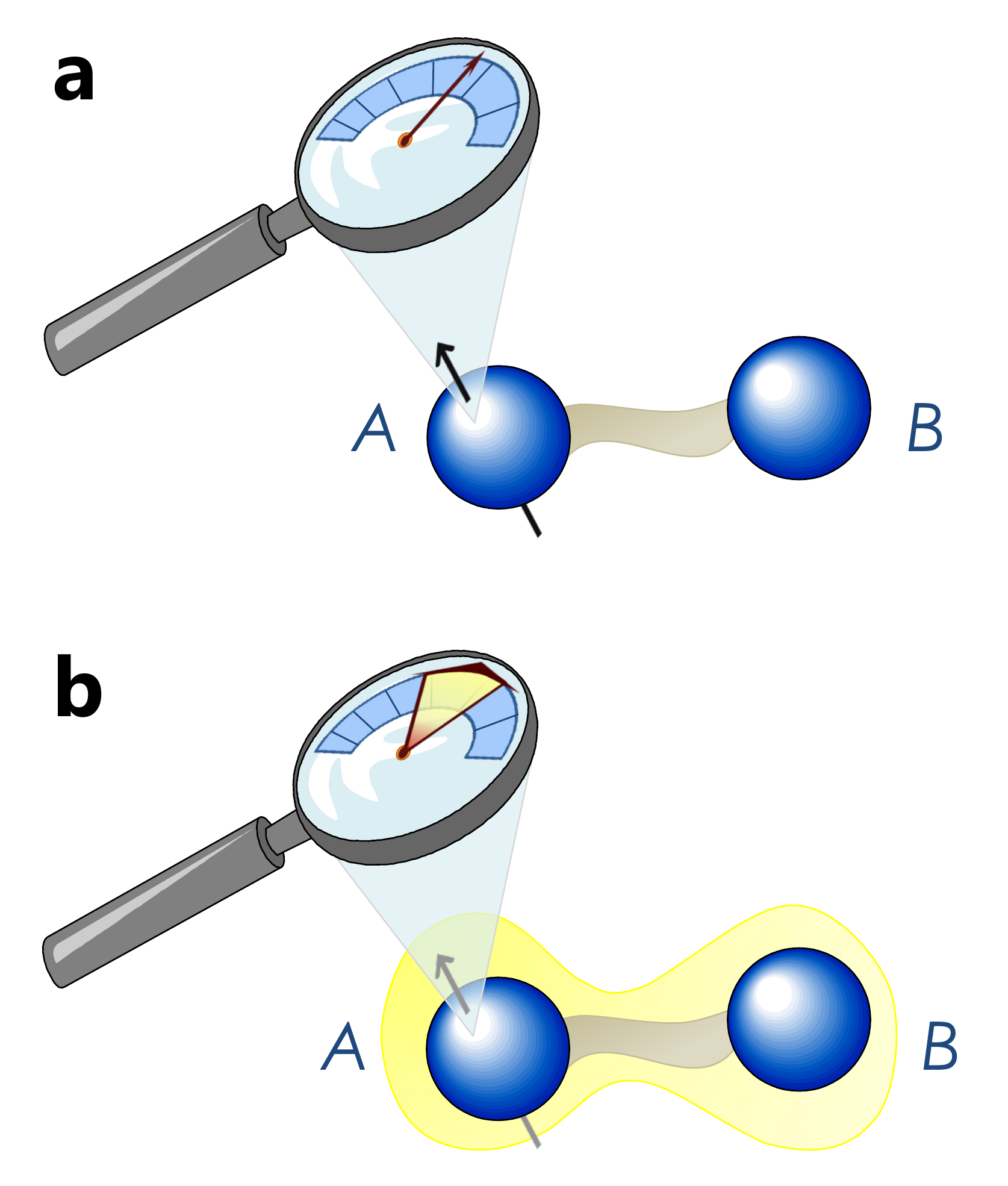}
\caption{(Color online) Quantum correlations trigger local quantum uncertainty. Let us consider a bipartite state $\rho$. An observer on subsystem $A$ is equipped with a {\em quantum meter}, a measurement device whose error bar shows the quantum uncertainty only (Note: in order to access such quantity, the measurement of other observables that are defined on the full bipartite system may be required, in a procedure similar to state tomography).
(a) If $\rho$ is uncorrelated or contains only classical correlations (brown shade), i.e. $\rho$ is of the form $\rho = \sum_i p_i \ket{i}\bra{i}_A \otimes {\sigma_i}_B$ (with $\{\ket{i}\}$ an orthonormal basis for $A$) \cite{OZ,HV,modirev}, the observer can measure at least one observable on $A$ without any intrinsic quantum uncertainty. (b) If $\rho$ contains a nonzero amount of quantum correlations (yellow shade), as quantified by entanglement for pure states \cite{entanglement} and quantum discord in general \cite{modirev}, any local measurement on $A$ is affected by quantum uncertainty. The minimum quantum uncertainty associated to a single measurement on subsystem $A$ can be used to quantify discord in the state $\rho$, as perceived by the observer on $A$.  In this Letter we adopt the Wigner-Yanase skew information \cite{wy} to measure the quantum uncertainty on local observables.}
\label{figurauno}
\end{figure}

Extending the argument to mixed states, one needs to filter out the uncertainty due to classical mixing, i.e., lack of knowledge of the state, in order to identify the genuinely quantum one.  We say that  an observable $K$ on the state $\rho$ is `quantum-certain' when the statistical error in its measurement is solely due to classical ignorance. By adopting a meaningful quantitative definition of quantum uncertainty, as detailed later, we find that $K$ is quantum-certain if and only if $\rho=\rho_K$, where $\rho_K$ is the density matrix of the state after the measurement of $K$. It follows that not only entangled states but also almost all (mixed) separable states
 \cite{ferraro} cannot admit any quantum-certain local observable. The only states left invariant by a local complete measurement are those described within classical probability theory \cite{perinotti}, i.e., embeddings of joint probability distributions. These are the states with zero {\it quantum discord} \cite{modirev,OZ,HV}.
The quantum uncertainty on local observables is then entwined to the notion of quantum discord (see Fig.~\ref{figurauno}), a form of nonclassical correlation which reduces to entanglement on pure states, and is currently subject to intense investigations for quantum computation and information processing \cite{dat,piani,np1,np2}.
In the following, an entire class of discordlike measures is defined, interpreted and analysed within the framework of local quantum uncertainty.

\begin{figure}[t]
\includegraphics[width=6.5cm]{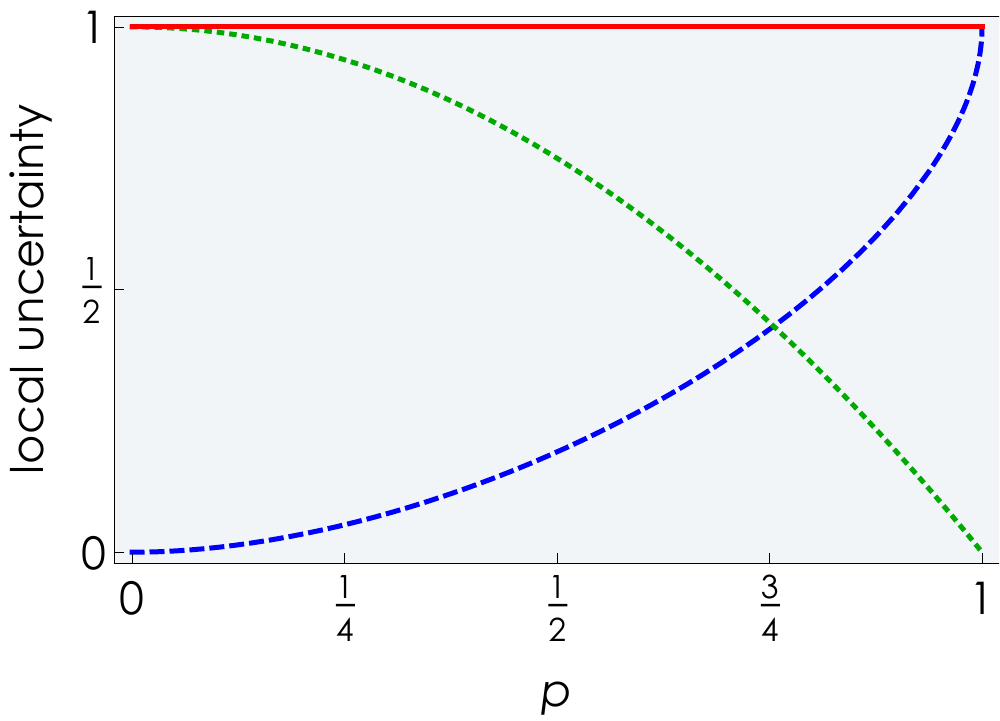}
\caption{(Color online)
The plot shows different contributions to the error bar of spin measurements on subsystem $A$ in a Werner state \cite{entanglement} $\rho = p |\phi^{+}\rangle\bra{\phi^{+}} + (1-p) \mathbb{I}/4, p\in [0,1],$ of two qubits $A$ and $B$. The red line is the variance $\text{Var}_\rho(\sigma_z^A)$ of the $\sigma_z^A$ operator, which amounts to the total statistical uncertainty.
The blue dashed curve represents the local quantum uncertainty ${\cal U}_A(\rho)$, which in this case is ${\cal I}(\rho,\sigma_z^A)$ (any local spin direction achieves the minimum for this class of states). The green dotted curve depicts the (normalized) linear entropy $S_L(\rho)=\frac43(1-\tr{}\{\rho^2\})$ of the global state $\rho$, which measures its mixedness. Notice that the Werner state is separable for $p\leq 1/3$ but it always contains discord for $p>0$.
 }\label{figurawerner}
\end{figure}

\noindent{\it Skew information and local quantum uncertainty.}---
 There are several ways to quantify the uncertainty on a measurement, and here we aim at extracting the truly quantum share. Entropic quantities or the variance, though employed extensively as indicators of uncertainty \cite{heis,unc,coles}, do not fit our purpose, since they are affected by the state mixedness. It has been proposed to isolate the quantum contribution to the total statistical error of a measurement as being due to the noncommutativity between state and observable: this may be reliably quantified via the {\it skew information} \cite{wy,luo1}
\begin{eqnarray}
{\cal I}(\rho,K)=-\frac 12 \text{Tr}\{[\rho^{\frac 12},K]^2\},\label{skewinfo}
\end{eqnarray}
introduced in \cite{wy} and employed for studies on uncertainty relations \cite{luo1}, quantum statistics and information geometry \cite{luo1,luo2,isola,brody,hansen}. Referring to \cite{wy} for the main properties of the skew information,  we recall the most relevant ones: it is nonnegative,  vanishing if and only if state and observable commute, and  is convex, that is, nonincreasing under classical mixing. Moreover, ${\cal I}(\rho,K)$ is always smaller than the
 variance of $K$, ${\cal I}(\rho,K) \leq \text{Var}_\rho(K)\equiv \med{{K}^{2}}_\rho-\med{K}_\rho^2$,
 with equality reached on pure states, where no classical ignorance occurs (see Fig.~\ref{figurawerner}).
Hence, we adopt the skew information as measure of quantum uncertainty and deliver a theoretical analysi in which we convey and discuss its operational interpretation.

 As a central concept in our analysis, we introduce the {\it local quantum uncertainty} (LQU) as the minimum  skew information achievable on a single local measurement. We remark that by `measurement' in the following we always refer to a complete von Neumann measurement. Let $\rho \equiv \rho_{AB}$ be the state of a bipartite system, and let $K^\Lambda=K^{\Lambda}_A \otimes  \mathbb{I}_{B}$ denote a local observable, with $K^{\Lambda}_A$ a Hermitian operator on $A$ with spectrum $\Lambda$. We require $\Lambda$ to be nondegenerate, which corresponds to maximally informative observables on $A$. The LQU with respect to subsystem $A$, optimized over all local observables on $A$ with nondegenerate spectrum $\Lambda$, is then
\begin{eqnarray}\label{gen}
{\cal U}^{\Lambda}_{A}(\rho)\equiv\min_{K^\Lambda} {\cal I}(\rho,K^\Lambda).\label{LQU}
\end{eqnarray}
Eq.~(\ref{gen}) defines a family of $\Lambda$-dependent quantities, one for each equivalence class of $\Lambda$-spectral local observables over which the minimum skew information is calculated. In practice, to evaluate the minimum in Eq.~\eqref{LQU}, it can be convenient to parametrize the observables on $A$ as  $K^{\Lambda}_A=V_A\text{diag}(\Lambda)V_A^\dagger$, where $V_A$ is varied over the special unitary group on $A$. In this representation, the (fixed) spectrum $\Lambda$ may be interpreted as a standard ``ruler", fixing the units as well as the scale of the measurement (that is, the separation between adjacent `ticks'), while $V_A$ defines the measurement basis that can be varied arbitrarily on the Hilbert space of $A$.

In the following, we prove some general qualitative properties of the $\Lambda$-dependent LQUs, which reveal their intrinsic connection with nonclassical correlations.

\noindent {\it A class of quantum correlations measures}.---
What characterizes a discordant state is, as anticipated, the non-existence of quantum-certain local observables. In fact, we find that each quantity ${\cal U}^{\Lambda}_{A}(\rho)$ defined in Eq.~(\ref{LQU}) is not only an indicator, but also a full fledged {\em measure} of bipartite quantum correlations (see Fig.~\ref{figurauno}) \cite{coles2}, i.e. it meets all the known {\it bona fide} criteria for a discordlike quantifier \cite{modirev}. Specifically, in the Supplemental Material \cite{supp} we prove that the $\Lambda$-dependent LQU (for any non-degenerate $\Lambda$) is invariant under local unitary operations, is nonincreasing under local operations on $B$, vanishes if and only if $\rho$ is a zero discord state with respect to measurements on $A$, and reduces to an entanglement monotone when $\rho$ is a pure state.

If we now specialize to the case of bipartite $2\times d$ systems, we further find that quantifying discord via the LQU is very advantageous in practice, compared to all the other measures proposed in the literature (which typically involve formidably hard optimizations not admitting a closed formula even for two-qubit states) \cite{modirev,geo}. Indeed, the minimization in  Eq.~(\ref{LQU}) can be expressed in closed form for arbitrary states $\rho_{AB}$ of a qubit-qudit system defined on $\mathbb C^2 \otimes \mathbb C^{d}$, so that ${\cal U}_A^\Lambda$ admits a {\it computable} closed formula. Moreover notice that, when $A$ is a qubit, all the $\Lambda$-dependent measures are equivalent up to a multiplication constant \cite{noteref}. We thus drop the superscript $\Lambda$ for brevity, and pick nondegenerate observables $K_A$ on the qubit $A$ of the form $K_A=V_A\sigma_{zA}V_A^\dagger=\vec n\cdot\vec\sigma_A$, with $|\vec n|=1$. This choice corresponds to a LQU normalized to unity for pure, maximally entangled states. Eq.~\eqref{LQU} can then be rewritten as the minimization of a quadratic form involving the unit vector $\vec n$, yielding simply
\begin{equation}
	{\cal U}_A(\rho_{AB})=1-\lambda_\text{max}\{W_{AB}\},\label{anale}
\end{equation}
where $\lambda_\text{max}$ denotes the maximum eigenvalue, and $W_{AB}$ is a $3\times3$ symmetric matrix whose elements are $$(W_{AB})_{ij}=\text{Tr}\left\{\rho_{AB}^{1/2}\,({\sigma_i}_A\otimes\mathbb I_B)\,\rho_{AB}^{1/2}\,({\sigma_j}_A\otimes\mathbb I_B)\right\},$$ with $i,j=x,y,z$.
It is easy to check that, for a pure state $\kebra{\psi_{AB}}{\psi_{AB}}$, Eq.~\eqref{anale} reduces to the linear entropy of entanglement, ${\cal U}_A(\kebra{\psi_{AB}}{\psi_{AB}})=2(1-\text{Tr}\;\rho_A^2)$, where $\rho_A$ is the marginal state of subsystem $A$. Qubit-qudit states represent a relevant class of states for applications in quantum information processing, and we present some pertinent examples in this Letter. The evaluation of the LQU for Werner states of two qubits is displayed in Fig.~\ref{figurawerner}. A case study of the discrete quantum computation with one bit (DQC1) model of quantum computation \cite{laf1}
is reported in the Supplemental Material \cite{supp}, showing that our measure (evaluated in the one versus $n$ qubits partition) exhibits the same scaling as the canonical entropic measure of discord \cite{OZ,dat}.
Beyond the practicality of having a closed formula, the approach adopted in this Letter provides in general a nice physical interpretation of discord as the minimum quantum contribution to the statistical variance associated to the measurement of local observables in correlated quantum systems.

Interestingly, the LQU in a general state $\rho_{AB}$ of a $\mathbb C^2 \otimes \mathbb C^{d}$ system, can be reinterpreted geometrically as the minimum squared Hellinger distance between $\rho_{AB}$ and the state after a least disturbing root-of-unity local unitary operation applied on the qubit $A$, in a spirit close to that adopted to define `geometric discords' based on other metrics \cite{modirev,stellar,faber,garbo,geo}. Let us recall that the squared Hellinger distance between density matrices $\rho$ and $\chi$ is defined as $D^2_H(\rho,\chi) = \frac12 \text{Tr}\{(\sqrt{\rho}-\sqrt{\chi})^2\}$ \cite{beng,luohell}. Observing that, for qubit $A$, any generic nondegenerate Hermitian observable $K_A=\vec n\cdot\vec\sigma_A$ is a root-of-unity unitary operation, which implies $K_A f(\rho_{AB}) K_A = f(K_A \rho_{AB} K_A)$ for any function $f$, we have
${\cal I}(\rho_{AB},K^A)=1-\text{Tr}\{\rho_{AB}^{\frac 12}K^A\rho_{AB}^{\frac 12}K^A\}=1-\text{Tr}\{\rho_{AB}^{\frac 12}(K^A\rho_{AB}K^A)^{\frac 12}\} = D_H(\rho_{AB}, K^A\rho_{AB}K^A)$; therefore, minimizing over the local observables $K^A=K_A \otimes \mathbb{I}_{B}$ yields the geometric interpretation of the LQU, analytically computed in Eq.~(\ref{anale}), in terms of Hellinger distance.  The study of further connections between uncertainty on a single local observable and geometric approaches to nonclassicality of correlations, possibly in larger and multipartite systems, opens an avenue for future investigations.

\begin{figure}[t]
\includegraphics[width=8.5cm]{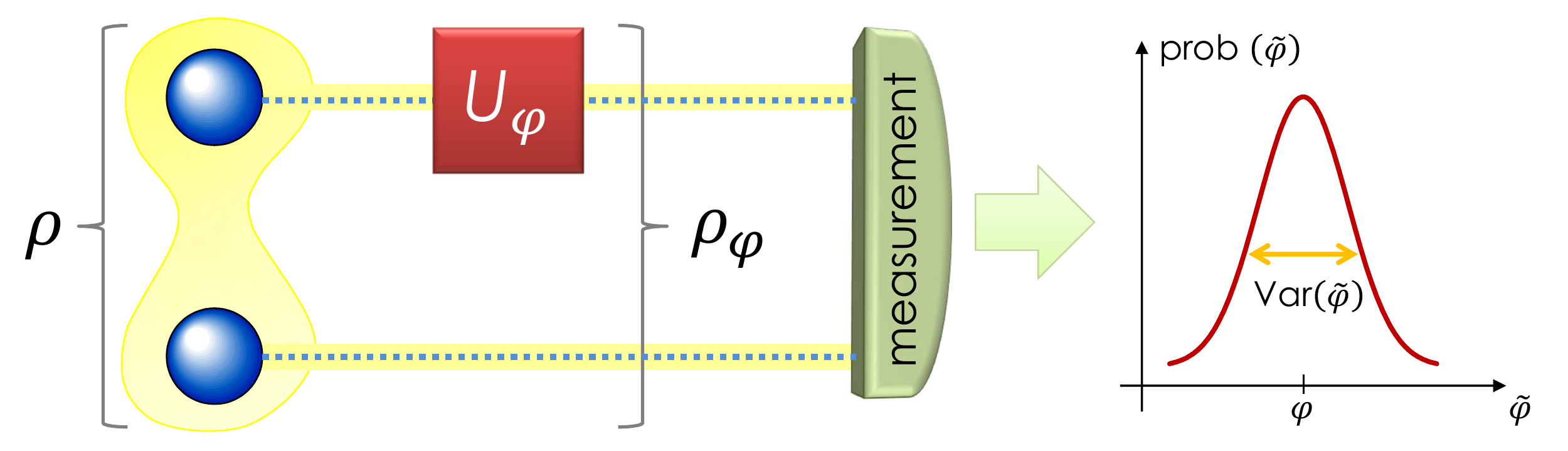}
\caption{(Color online)  Quantum correlations-assisted parameter estimation. A probe state $\rho$ of a bipartite system $AB$ is prepared, and a local unitary transformation depending on an unobservable parameter $\varphi$ acts on subsystem $A$, transforming the global state into $\rho_{\varphi}$. By means of a suitable measurement at the output one can construct an (unbiased) estimator $\tilde{\varphi}$ for $\varphi$. The quality of the estimation strategy is benchmarked by the variance of the estimator. For a given probe state $\rho$, the optimal measurement at the output  returns
an estimator $\tilde{\varphi}_{\rm best}$ for $\varphi$ with the minimum allowed variance given by the inverse of the QFI ${\cal F}(\rho_\varphi)$, according to the quantum Cram\'er-Rao bound \cite{quantumcramer}.
In the prototypical case of optical phase estimation,  the present scheme corresponds to a Mach-Zender interferometer.
Restricting to pure inputs, research in quantum metrology \cite{metrology} has shown that in this case entangled probes allow to beat the shot noise limit ${\cal F} \propto n$ ($n$ being the input mean photon number) and reach ideally the Heisenberg scaling ${\cal F} \propto n^2$. However, recent investigations have revealed how in presence of realistic imperfections the achieved precision quickly degrades to the shot noise level \cite{rafau,madalin,davido}.  For mixed bipartite probes, we show that the QFI is bounded from below by the amount of quantum correlations in the probe state $\rho$ as quantified by the LQU.}
\label{metrofiga}
\end{figure}

\noindent{\it Applications to quantum metrology.}--- We now discuss the operative role that discord, as quantified by the LQU, can play in the paradigmatic scenario of phase estimation in quantum metrology \cite{metrology}. We focus here on an `interferometric' setup employing bipartite probe states, as sketched in Fig.~\ref{metrofiga}.

 Given a (generally mixed) bipartite state $\rho$ used as a probe, subsystem $A$ undergoes a unitary transformation (specifically, a phase shift) so that the global state changes to $\rho_\varphi =  U_\varphi \rho U^\dagger_\varphi$, where $U_\varphi = {\rm e}^{-i \varphi H_A}$, with $H_A$ a local Hamiltonian on $A$, which we assume to have a nondegenerate spectrum $\bar\Lambda$. The goal is to estimate the unobservable parameter $\varphi$. The protocol, which has wide-reaching applications, from gravitometry to sensing technologies \cite{metrology,paris}, can be optimized by picking the best probe state $\rho$ and the most informative measurement at the output. It is known that the latter optimization can be solved in general by choosing, for any probe state $\rho$,  the measurement strategy which saturates asymptotically the quantum Cram\'er-Rao bound, $\text{Var}(\tilde{\varphi}) \geq 1/[\nu {\cal F}(\rho_\varphi)]$ \cite{quantumcramer}, where the quantum Fisher information (QFI) ${\cal F}(\rho_\varphi)$ sets then the precision of the optimal estimation, and $\nu$ denotes the number of times the experiment is repeated ($\nu\gg1$ is assumed). We will denote by $\tilde{\varphi}_{\rm best}$ the estimator obtained from the optimal measurement strategy, so that $\text{Var}(\tilde{\varphi}_{\rm best}) = 1/[\nu {\cal F}(\rho_\varphi)]$. Recall that the QFI can be written as \cite{paris,petz} ${\cal F}(\rho_\varphi) = \text{Tr}\{\rho_\varphi L^2_\varphi\}$, with $L_\varphi$ being the symmetric logarithmic derivative defined implicitly by $2 \partial_\varphi \rho_\varphi = L_\varphi \rho_\varphi + \rho_\varphi L_\varphi$.

We focus therefore on the optimization of the input state. In practical conditions, e.g.~when the engineering of the probe states occurs within a thermal environment or with a reduced degree of control, it may not be possible to avoid some degree of mixing in the prepared probe states. It is then of fundamental and practical importance to investigate the achievable precision when the phase estimation is performed within specific noisy settings \cite{rafau,madalin,davido}.
Here we assess whether and how quantum correlations in the (generally mixed) state $\rho$ play a role in determining the sensitivity of the estimation. Notice that the remaining steps of the estimation process are assumed to be noiseless (the unknown transformation $U_\varphi$ is unitary and the output measurement is the ideal one defined above). The key observation stems from the relation between the Wigner-Yanase and the Fisher metrics \cite{petz}, which implies that the skew information of the Hamiltonian is majorized by the QFI \cite{luo1,luoter}. As $H_A$ is  not necessarily the most certain local observable with spectrum $\bar \Lambda$, the $\bar \Lambda$-LQU itself fixes a lower bound to the QFI:
\begin{equation}\label{premestruo}{\cal U}_{A}^{\bar\Lambda}(\rho) \leq {\cal I}(\rho,H_A)= {\cal I}(\rho_{\varphi},H_A) \leq \hbox{$\frac14$} {\cal F}(\rho_\varphi).
\end{equation}
Then, for probe states with any nonzero amount of discord, and for $\nu \gg 1$ repetitions of the experiment, the optimal detection strategy which asymptotically saturates the quantum Cram\'er-Rao bound produces an estimator $\tilde{\varphi}_{\rm best}$ with necessarily limited variance, scaling as
   \begin{eqnarray}
 \text{Var}(\tilde{\varphi}_{\rm best})=\frac{1}{\nu {\cal F}(\rho_\varphi)} \leq \frac{1}{4 \nu {\cal U}_{A}^{\bar\Lambda}(\rho)}.\label{metrosexual}
 \end{eqnarray}
Hence, we established on rigorous footings that the quantum correlations measured by LQU, though not necessary \cite{pezze,modix,boixo}, are a sufficient resource to ensure a guaranteed
upper bound on the smallest possible variance with which a phase $\varphi$ can be
measured with mixed probes.

We now provide a simple example to clarify the above general discussion.
Suppose system $A$ is a spin-$j$ particle undergoing a phase rotation $U_\varphi=\exp(-i\varphi J_z)$, where $J_z$ is the third spin component, and $\varphi$ the phase to be estimated. In this case the estimation precision is bounded by the so-called {\it Heisenberg limit} ${\cal F}_\text{max}=4j^2$ \cite{vaneph,metrology}. A typical scheme achieving this limit can be outlined as follows. Assume that system $B$ is simply a qubit with states $|0\rangle_B,|1\rangle_B$. The $AB$ system is initially prepared in the product state $\ket{j}_A\ket{+}_B$, where $\ket m_A$ are the eigenstates of $J_z$ with eigenvalues $m=-j,-j+1,...,j$,  and $\ket{\pm}_B=\tfrac1{\sqrt2}(\ket0_B\pm\ket1_B)$.
Then, a `control-flip' operation $\propto\exp(i\pi J_{xA}\kebra{1}{1}_B)$ is applied, so that the system evolves to $\ket{\psi}_{AB}=\tfrac1{\sqrt2}(\ket j_A\ket0_B+\ket{\!-\! j}_A\ket 1_B)$. One can see that the entangled state $\ket{\psi}_{AB}$ used as a probe
achieves the Heisenberg limit. Our general treatment allows us to study quantitatively the effect of noise on the estimation power of the bipartite state $\ket\psi_{AB}$. Suppose now that the probe state, ideally $\ket{\psi}_{AB}$, is prepared in a noisy environment, which induces partial dephasing in the basis $\ket{m}_A$. Then, our probe state is in general given by
$\!\rho_{AB}\!=\!\tfrac12\left[\kebra{j,0}{j,0}+\kebra{-\!j,1}{-\!j,1}+r\left(\kebra{j,0}{-\!j,1}\!+\!\text{H.c.}\right)\right],$
where $0\!\leq\! r\!\leq\!1$ quantifies the degree of residual coherence, and $\ket{m,\phi}\equiv\ket{m}_A\ket{\phi}_B$. As this is effectively a $2$-qubit state, we can restrict our analysis to a truncated $2\times2$ Hilbert space. Here the restriction of $J_z$ has the spectrum $\bar\Lambda=(-j,j)$. We can thus calculate the $\bar\Lambda$-LQU in this effective $2\times2$ Hilbert space, obtaining ${\cal U}_A^{\bar \Lambda}=j^2(1\!-\!\sqrt{1\!-\!r^2})$. For any $j$, notice that the discord is a monotonically increasing function of the coherence $r$. Hence, from Eq.~(\ref{premestruo}) one has
${\cal F}(\rho_{AB}^{\varphi})\geq4{\cal U}_A^{\bar \Lambda}(\rho_{AB})=4j^2(1\!-\!\sqrt{1\!-\!r^2}).$ As the spin number $j$ is increasing, this guarantees that the classical scaling ${\cal F}\sim 2j$ (i.e. the so-called {\it shot noise limit} \cite{metrology}) can still be beaten provided that $r\gtrsim1/\sqrt j$.

The connection between the LQU and the sensitivity of parameter estimation can also be appreciated in more abstract geometrical terms, without the need for invoking the Fisher information. As shown by Brody \cite{brody}, the skew information ${\cal I}(\rho_{\varphi},H_A)$ of the Hamiltonian $H_A$ determines the squared speed of evolution of the density matrix $\rho$ under the unitary $U_\varphi = {\rm e}^{-i \varphi H_A}$. This provides another geometric interpretation for the LQU: The observable $K_A$ which achieves the minimum in Eq.~{\ref{LQU}} is the local observable with the property that the resulting local unitary operation ${\rm e}^{-i \varphi K_A}$ makes the given state $\rho$ of the whole system evolve as slowly as possible (the observable $K_A$  is the least disturbing in this specific sense). Since a higher speed of state evolution under a change in the parameter $\varphi$ means a higher sensitivity of the given probe state to the estimation of the parameter, our result can be interpreted as follows: The amount of discord (LQU) in a mixed correlated probe state $\rho$ used for estimation of a parameter $\varphi$ bounds from below the squared speed of evolution of the state under any local Hamiltonian evolution ${\rm e}^{-i \varphi H_A}$, hence the sensitivity of the given probe state $\rho$ to a variation of $\varphi$, which is a general measure of precision for the considered metrological task.

\noindent{\it Conclusions.}---
In this Letter we studied the quantum uncertainty on single observables. The exploration of this concept allowed us to define and investigate a class of measures of bipartite quantum correlations of the discord type \cite{modirev}, which are physically insightful and mathematically rigorous. In particular, for qubit-qudit states a unique measure is defined (up to normalization), and it is computable in closed form. Quantum correlations, in the form known as quantum discord \cite{OZ,HV}, manifest in the fact that any single local observable displays an intrinsic quantum uncertainty. Discord in mixed probe states, measured by the local quantum uncertainty, is further proven to guarantee a minimum sensitivity in the protocol of optimal phase estimation \cite{metrology}.
We believe worthwhile to substantiate in future work the promising uncovered connections between quantum mechanics, information geometry  and complexity science \cite{quantumcramer,hansen,gu} by addressing the role of quantum uncertainty, in particular induced by quantum correlations, in such contexts.\\
\noindent {\it Note added in proof.} Very recently, an alternative measure of discord based on the quantum Hellinger distance has been proposed in Ref.~\cite{hell2013}.

\noindent {\it Acknowledgments.}--- D.G. and G.A. thank V.P. Belavkin {\it in memoriam}  for having shared quantum bits of his awesome talent and his encyclopedic knowledge of the laws of Nature.
We warmly thank L. Correa for extensively commenting on the manuscript, and  A. Balyuk, M. Ahmadi, D. Brody, J. Calsamiglia, F. Ciccarello, P. J. Coles,  A. Datta, M. G. Genoni, V. Giovannetti, S. Girolami, M. Guta, F. Illuminati, M. S. Kim, S. Luo, L. Maccone, S. Pascazio, M. Piani, T. Rudolph for fruitful discussions. We acknowledge financial support from Universitas 21, the University of Nottingham, the EPSRC Research Development Fund
(Grant No. PP-0313/36), and the Qatar National Research Fund (Grant No. NPRP 4-554-1-084).

\clearpage

\setcounter{page}{1}

\appendix*
\begin{widetext}

\section{Supplemental Material}
\begin{center}
{\large{\bf {Characterizing Nonclassical Correlations via Local Quantum Uncertainty}}}

\quad \\

{\normalsize Davide Girolami, Tommaso Tufarelli, and Gerardo Adesso}

\end{center}

\setcounter{equation}{0}

\bigskip

\quad\\

\begin{center}
{\bf Proof of the properties of LQU}
\end{center}

 We refer to \cite{wy,luo1,luo2} for a summary of the relevant properties of the skew information which constitute the main ingredients of the proofs. In a bipartite system $AB$, classically correlated states $\rho_c$ with respect to measurements on $A$, also known as $A$-classically correlated states or classical-quantum states, are states with zero quantum discord on $A$.  For these states there exists at least one set of projectors $\{\Pi_i=\Pi^A_i\otimes \mathbb{I}^B\}$ such that $\rho_c=\sum_i \Pi_i \rho_c\Pi_i$. The $A$-classically correlated states take in general the form $\rho_c=\sum_i p_{i} |i\rangle\langle i|_A\otimes {\tau_i}_B$ with $\{\ket{i}\}$ denoting an orthonormal basis for subsystem $A$.

To prove that $A$-classically correlated states have vanishing LQU ${\cal U}^{\Lambda}_A$, it is sufficient to define the observable $K^{\Pi}=K_{A}^{\Pi}\otimes \mathbb{I}_B$ where $K_A^{\Pi}$ is diagonal in the basis defined by $\{\Pi^A_i\}$, to obtain $[\rho_c, K_{\Pi} ]=0$ which means ${\cal U}_A^{\Lambda}(\rho_c)={\cal I}(\rho_c, K_{\Pi})=0$. On the other hand, a vanishing LQU ensures the existence of a local observable $\tilde{K}^A$ such that  ${\cal I}(\rho, \tilde{K}^A)=0$. Hence $\tilde K^A$ commutes with the density matrix, and we can diagonalize them simultaneously. Since the observable is assumed nondegenerate, its eigenvectors define a {\it unique} basis on $A$ (up to phases), say $\{\ket{k_i}\}$. Then, an eigenvector basis for $\tilde{K}_A$ will be simply $\{|k_{i}\rangle_A\otimes |\phi_{ij}\rangle_{B}\}$, and the state must necessarily be of the form $\rho_{K_A}=\sum_i p_{ij} |k_{i}\rangle\langle k_{i}|_A\otimes |\phi_{ij}\rangle\langle \phi_{ij}|_B$, which is a zero discord state. This proves that ${\cal U}^{\Lambda}_A(\rho)$ vanishes if and only if $\rho$ is an $A$-classically correlated state.

Let us now show that the LQU is invariant under local unitary transformations. We have\\
${\cal U}^{\Lambda}_A\left((U_A\otimes U_B) \rho (U_A\otimes U_B)^\dagger\right)$ \\
$=\min_{K^A} {\cal I}\left((U_A\otimes U_B)\rho(U_A\otimes U_B)^\dagger, K_A \otimes \mathbb I_B\right)$ \\
$=\min_{K^A} {\cal I}\left(\rho, (U_A\otimes U_B)^\dagger (K_A \otimes \mathbb I_B)(U_A\otimes U_B)\right)$ \\
$= \min_{K^A} {\cal I}\left(\rho, (U_A^\dagger K_A U_A)\otimes  \mathbb I_B\right) = {\cal U}^{\Lambda}_A(\rho)$,\\
 as minimizing over the local observables $K^A$ is obviously equivalent to do it over the ones rotated by $U_A$.

We then note that the skew information ${\cal I}(\rho, K^A)$ is contractive under completely positive and trace-preserving maps $\Phi_B$ on $B$, ${\cal I}(\rho,K_A \otimes \mathbb I_B) \geq {\cal I}\big((\mathbb{I}_A\otimes \Phi_B)\rho,K_A \otimes \mathbb I_B\big)$. Consequently, the LQU inherits this property. Denoting as $\tilde{K}_A$ the most certain observable for $\rho$, we have
${\cal U}^{\Lambda}_A(\rho)={\cal I}(\rho, \tilde{K}_A \otimes \mathbb I_B) \geq {\cal I}\left((\mathbb{I}_A\otimes \Phi_B)\rho, \tilde{K}_A \otimes \mathbb I_B\right)\geq  {\cal U}^{\Lambda}_A\big((\mathbb{I}_A\otimes \Phi_B)\rho\big)$.

Finally, for pure states $\rho=\ket{\psi}\!\bra{\psi}$, the LQU reduces to the variance of $K^A$ minimized over all local observables $K^A$. In the next section we present a proof (which can be of independent interest) that such a quantity  decreases monotonically under local operations and classical communication, so that the LQU, alias minimal local variance, reduces to an entanglement measure on pure states. \\

\begin{center}
{\bf Proof of LOCC monotonicity of LQU for pure states}
\end{center}
\begin{Lemma}\label{permutanda}
Consider a $N$-dimensional density matrix $\rho$, and the set $\{K\}$ of all observables with fixed spectrum $\Lambda=(\lambda_1,...,\lambda_N)$. Then, the variance $\text{Var}_\rho(K)\equiv\mathcal{V}(\rho,K)=\tr{} \{\rho K^2\}-\tr{}\{\rho K\}^2$ is minimised by an observable $K_0$ commuting with $\rho$.
\begin{proof}
Working in the eigenbasis of the density matrix, one has the representation $\rho={\rm diag}(p_1,...,p_N)$. An observable in the considered set can the be written as $K=V{\rm diag}(\lambda_1,...,\lambda_N)V^\dagger$, where $V$ is a unitary transformation. The variance of $K$ on the state $\rho$ reads ($V_{ij}\equiv\bra i V\ket j$)
\begin{equation}
\mathcal{V}(\rho,K)=\sum_{i,j}p_i\lambda_j^2|V_{ij}|^2-\left(\sum_{i,j}p_i\lambda_j|V_{ij}|^2\right)^2\equiv\tr{}\{PB\}-[\tr{}\{QB\}]^2\qquad P_{ij}\equiv p_i\lambda_j^2,\;Q_{ij}\equiv p_i\lambda_j,\;B_{ij}\equiv|V_{ij}|^2.\label{boh}
\end{equation}
Note that $B$ is a unistochastic matrix, and in fact, any unistochastic matrix is expressible as $B_{ij}=|V_{ij}|^2$ for some unitary $V$. Hence, the problem of minimizing the variance can be equivalently formulated as a minimization of the right hand side of Eq.~\eqref{boh} over the set of unistochastic matrices. Since every unistochastic matrix is also bistochastic (but not vice-versa), one has, in general
\begin{equation}
\min_{\{K\}}\mathcal V(\rho,K)\geq \min_{B\in\mathcal B}\left[\tr{}\{PB\}-[\tr{}\{QB\}]^2\right],\label{varbound}
\end{equation}
where $\mathcal B$ is the set of all $N\times N$ bistochastic matrices. One can now exploit the Birkhoff-von Neumann theorem, and express a generic bistochastic matrix as a convex sum of permutations of the form $B=\sum_k q_k S_k$, where the $q_k$'s are probabilities and $\{S_k\}$ is the set of permutation matrices in dimension $N$, which has $N!$ elements. Then,
\begin{align}
\min_{B\in\mathcal B}\left[\tr{}\{PB\}-[\tr{}\{QB\}]^2\right]&=\min_{\{q_k\}}\left[\sum_kq_k\tr{}\{PS_k\}-\left(\sum_kq_k\tr{}\{QS_k\}\right)^2\right]\geq\min_{\{q_k\}}\sum_kq_k\left[\tr{}\{PS_k\}-\left[\tr{}\{QS_k\}\right]^2\right]\nonumber\\
&\geq\sum_kq_k\left[\tr{}\{PS_\text{min}\}-\left[\tr{}\{QS_\text{min}\}\right]^2\right]=\tr{}\{P S_\text{min}\}-[\tr{}\{QS_\text{min}\}]^2,\
\end{align}
where we have exploted the convexity of the square, and $S_\text{min}$ is a particular permutation that minimises the expression $\tr{}\{PS_k\}-\left[\tr{}\{QS_k\}\right]$. Such minimizing permutation can always be found since $\{S_k\}$ is a finite set. Noting that permutations are also unistochastic matrices, the above steps imply that the equality sign in Eq.~\eqref{varbound} can be always achieved:
\begin{equation}
\min_{\{K\}}\mathcal V(\rho,K)=\tr{}\{P S_\text{min}\}-[\tr{}\{QS_\text{min}\}]^2=\sum_i p_i\lambda_{\mathcal{P}(i)}^2-\left(\sum_ip_i\lambda_{\mathcal{P}(i)}\right)^2,
\end{equation}
where $\mathcal P$ indicates the permutation of the indices associated to the matrix $S_\text{min}$. This implies that the variance is minimised by an observable of the form $K_0={\rm diag}(\lambda_{\mathcal P(1)},...,\lambda_{\mathcal P(N)})$, which clearly commutes with $\rho.$
\end{proof}
\end{Lemma}
 \begin{Lemma}\label{lem1}
Let $d_{A,B}\equiv\text{dim}(\mathcal H_{A,B})$. Suppose that $d_A\leq d_B$. Under local operations on subsystem $A$, a globally pure state $|\psi\rangle$ evolves within a subspace $\mathcal H_A\otimes\tilde{\mathcal H}_B$, where $\tilde{\mathcal H}_B$ is a $d_A$-dimensional subspace of $\mathcal H_B$.
\begin{proof} We can suppose that $|\psi\rangle$ is in Schmidt form:
\begin{equation}
|\psi\rangle=\sum_j^{d_A}c_i|i_A\rangle| i_B\rangle,\label{schmidt}
\end{equation} Clearly,
$|\psi\rangle\in\mathcal H_A\otimes\tilde{\mathcal H}_B$, where $\tilde{\mathcal H}_B$ is spanned by the $d_A$ orthonormal vectors $\{|i_B\rangle\}$. A local operation on $A$ is described via Kraus operators of the form $M^A=M_A\otimes\mathbb I_B$. Applying the operator on the state, one has:
\begin{equation}
	M^A|\psi\rangle=\sum_j^{d_A}c_i(M_A|i_A\rangle )| i_B\rangle,\label{subvec}
\end{equation}
which is still a vector with support in $\mathcal H_A\otimes\tilde{\mathcal H}_B$.
\end{proof}
\end{Lemma}
\begin{Corollary}
When applying operations on $A$ to a pure state, we can suppose $d_A\geq d_B$. A proof of monotonicity in this particular case will then be sufficient.
\end{Corollary}
\begin{Lemma}\label{lem2}
Suppose $d_A\geq d_B$, and the (non-degenerate) spectrum of the $A$-observables is fixed as $\Lambda(K_A)=\{\lambda_{1},...,\lambda_{d_A}\}$.
One has,
\begin{eqnarray}
{\cal U}^{\Lambda}_A(|\psi\rangle\langle\psi|)=\min_{K_B\in\mathcal K_B}{\cal I}(|\psi\rangle\langle\psi|,\mathbb {I}_A\otimes K_B),\label{l2}
\end{eqnarray}
where $\mathcal K_B$ is the set of B-observables whose $d_B$ eigenvalues are non degenerate and are a {\it subset} of $\Lambda(K_A)$: $\Lambda(K_B)=\{\mu_1,...,\mu_{d_B}|\mu_j\in\Lambda(K_A),\mu_i\neq\mu_j (i\neq j)\}$.
\begin{proof} We start by noting that, in general, ${\cal U}^{\Lambda}_A(|\psi\rangle\langle\psi|)\leq\min_{K_B\in\mathcal K_B}{\cal I}(|\psi\rangle\langle\psi|,\mathbb I_A\otimes K_B)$. In fact, by rotating $|\psi\rangle$ to the Schmidt form, we see that the variance of any observable $K_B\in\mathcal K_B$ is achieved by an operator $K_A$ on $A$. Given $K^B$ such that $K^B|\psi\rangle=\sum_{ij}c_i(K_B)_{ij}|i_A\rangle| j_B\rangle$, it is sufficient to choose $K^A$ such that $K^A|\psi\rangle=\sum_{ij}c_i(K_A)_{ij}|j_A\rangle| i_B\rangle=\sum_{ij}c_i(K_A)_{ij}| i_B\rangle| j_A\rangle$. The two operators clearly yield the same variance, since the labels $A,B$ do not affect its calculation. Note that it is always possible to pick $K^A$ in the above form since the operators on $A$ restricted to a $d_B$-dimensional subspace can assume the same form as any operator in $\mathcal K_B$.

We now show that the inequality ${\cal U}^{\Lambda}_A(|\psi\rangle\langle\psi|)\geq\min_{K_B\in\mathcal K_B}{\cal I}(|\psi\rangle\langle\psi|,\mathbb I_A\otimes K_B)$ is also verified, hence equality must hold. The most certain observable on $A$ has to commute with the reduced state $\rho_A$ (Lemma ~\ref{permutanda}). Hence, if the latter has eigenvalues $p_j$, $j\leq d_B$, there is an appropriate permutation $\mathcal P$ such that:
\begin{equation}
{\cal U}^{\Lambda}_A(\psi)=\sum_{j=1}^{d_B}p_j(\lambda_{\mathcal P(j)})^2-\left(\sum_{j=1}^{d_B}p_j\lambda_{\mathcal P(j)}\right)^2={\cal I}(|\psi\rangle\langle\psi|,\mathbb I_A\otimes \tilde K_B).
\end{equation}
The latter equality is obtained by choosing $\tilde K_B$ diagonal in the same basis as $\rho_B$, with eigenvalues $\mu_j=\lambda_{\mathcal P(j)}$, and by noting that $\rho_A$ and $\rho_B$ have the same eigenvalues.
\end{proof}
\end{Lemma}
\begin{Theorem}
The LQU is an entanglement monotone for pure states.
\begin{proof}
By Lemma~\ref{lem1}, we can suppose $d_A\geq d_B$. We already have invariance under local unitaries and contractivity under local operations on $B$ (see Appendix in the main text). To complete the proof we need to prove that, on average, the LQU of $\ket{\psi}$ cannot be increased under operations on $A$. Let $\{ M_i^A \}$ be the Kraus operators on Alice: $\sum_i M_{i}^{A \dagger} M_{i}^A=\mathbb I$. The output ensemble is given by $\{p_i,|\phi_i\rangle\}$, where
\begin{equation}
\sqrt{p_i}|\phi_i\rangle=M_i^A|\psi\rangle.
\end{equation}
We want to demonstrate that $\sum_ip_i{\cal U}^{\Lambda}_A(|\psi\rangle\langle\psi|)\leq{\cal U}^{\Lambda}_A(|\psi\rangle\langle\psi|)$. Suppose that $K_0\in\mathcal K_B$ is such that ${\cal U}^{\Lambda}_A(|\psi\rangle\langle\psi|)={\cal I}(|\psi\rangle\langle\psi|,\mathbb I_A\otimes K_0)$, as given by Lemma~\ref{lem2}.
\begin{eqnarray}
&\sum_ip_i{\cal U}^{\Lambda}_A(|\phi_i\rangle\langle\phi_i|)=\sum_ip_i\min_{K_i\in\mathcal K_B}\mathcal I(|\phi_i\rangle\langle\phi_i|,\mathbb{I}_A\otimes K_i)\leq\sum_ip_i\mathcal I(|\phi_i\rangle\langle\phi_i|,\mathbb{I}_A\otimes K_0)\nonumber\\
&=\sum_ip_i\mathcal V(|\phi_i\rangle\langle\phi_i|,\mathbb{I}_A\otimes K_0)\leq \mathcal{V}\left(\sum_ip_i|\phi_i\rangle\langle\phi_i|,\mathbb{I}_A\otimes K_0\right)\nonumber\\
&=\sum_ip_i\langle\phi_i|\mathbb{I}_A\otimes K_0^ 2|\phi_i\rangle-\left(\sum_ip_i\langle\phi_i|\mathbb{I}_A\otimes K_0|\phi_i\rangle\right)^2\nonumber\\
&=\sum_i\langle\psi| M^A_i(\mathbb{I}_A\otimes K_0^2)M^{A \dagger}_i|\psi\rangle-\left(\sum_i\langle\psi|M^A_i(\mathbb{I}_A\otimes K_0)M^{A \dagger}_i|\psi\rangle\right)^2\nonumber\\
&=\langle\psi|\sum_iM_{iA}^{\dagger} M_{iA}^{}\otimes K_0^2|\psi\rangle-\left(\langle\psi|\sum_iM_{iA}^{\dagger} M_{iA}^{}\otimes K_0|\psi\rangle\right)^2\nonumber\\
&=\langle\psi|\mathbb{I}_A\otimes K_0^2|\psi\rangle-\left(\langle\psi|\mathbb{I}_A\otimes K_0|\psi\rangle\right)^2={\cal I}(|\psi\rangle\langle\psi|,\mathbb I_A\otimes K_0)={\cal U}^{\Lambda}_A(|\psi\rangle\langle\psi|).
\end{eqnarray}
 In the first line, we used Lemma ~\ref{lem2}. In the second line, we have used that the variance is concave as a function of the state.\\
\end{proof}
\end{Theorem}

 \ \\

\begin{center}
{\bf Example: LQU in the DQC1 model}\\
\end{center}

An interesting case study concerns the final state of the  DQC1 (Discrete Quantum Computation with One bit) model, a protocol designed for estimating the trace of a unitary matrix, say $U$, applied on a $n$-qubit register \cite{laf1}.  Discordlike correlations, but vanishing entanglement, are created between an ancillary qubit and the register   in the output state \cite{dat}. The ancilla $A$, in a state with arbitrary polarization $\mu$, say $\rho^{in}_A=\frac12(\mathbb{I}_2+\mu \sigma_3)$ , and the register $B$, in a $n$-qubit maximally mixed state, i.e.,  $\rho^B_{in}=\frac1{2^n}\mathbb{I}_{n}$,  are initially uncorrelated:  $\rho^{in}=\rho^{in}_A\otimes\rho^{in}_B$. The protocol returns the final state
\begin{eqnarray}
\rho^{out}&=&\frac1{2^{n+1}}\left(
\begin{array}{c|c}
 \mathbb{I}_{n} & \mu U^{\dagger}  \\ \hline
 \mu U &   \mathbb{I}_{n}\\
\end{array}
\right).\label{rhoout}
  \end{eqnarray}

Measuring the ancilla polarization in the output state yields an estimation of the trace of the unitary matrix: $\langle\sigma_1\rangle_{\rho_A^{out}}=\text{Re}\left[\text{Tr}[U]\right], \langle\sigma_2\rangle_{\rho_A^{out}}=\text{Im}\left[\text{Tr}[U]\right]$. For `typical' unitaries in high dimensions (which have approximately zero trace \cite{dat}), the entanglement between the ancilla and the $n$-qubit register is always negligible. On the other hand, we find the following.
\begin{Rem}
 The local quantum uncertainty calculated via Eq.~\eqref{anale} yields:
 \begin{eqnarray}
 {\cal U}_{A}(\rho^{\text{out}})=\frac{1}{2}\left(1-\sqrt{1-\mu ^2}\right).\label{dqc1}
 \end{eqnarray}
\begin{proof}
 We choose the basis $\{\ket{k}\}$ on $B$ which diagonalizes $U$: $U\ket{k}={\rm e}^{-i\varphi_k}\ket k$. We may then rewrite Eq.~\eqref{rhoout} as $\rho^{out}=2^{-n}\sum_k\rho_k\otimes\kebra{k}{k}$, where $\rho_k=1/2(\mathbb I_A+\vec\mu_k\cdot\vec\sigma)$ and $\vec \mu_k=\mu(\cos{\varphi_k},\sin{\varphi_k},0)$. The square root of the density matrix can then be expressed as $\sqrt{\rho^{out}}=2^{-n/2}\sum_k r_k\otimes\kebra{k}{k}$, where $r_k=2^{-1/2}( v_0 \mathbb I_A+\vec v_k\cdot\vec \sigma)$, where $\vec v_k=v(\cos{\varphi_k},\sin{\varphi_k},0)$ and the pair $v_0,v$ verify $v_0^2+v^2=1$ and $2v_0v=\mu$.  Both $v_0$ and $v\equiv|\vec v_k|$ do not depend on $k$, while $\vec v_k$ does. The elements of the matrix $W_{AB}$ are then given by
\begin{align}
	(W_{AB})_{ij}&=\frac{1}{2^{n}}\sum_{k}\text{Tr}\{r_k\sigma_i r_k\sigma_j\}\nonumber\\
	&=v_0^2\delta_{ij}+2^{-(n+1)}\sum_{k,l,m}(\vec v_k)_l(\vec v_k)_m\text{Tr}\{\sigma_i\sigma_l\sigma_j\sigma_m\}.
\end{align}
Now, we see that $\text{Tr}\{\sigma_i\sigma_l\sigma_j\sigma_m\}=2(\delta_{il}\delta_{jm}-\delta_{ij}\delta_{lm}+\delta_{im}\delta_{jl})$. Hence,
\begin{align}
	(W_{AB})_{ij}&=(v_0^2-v^2)\delta_{ij}+\frac{2}{2^{n}}\sum_{k}(\vec v_k)_i(\vec v_k)_j.\label{nastysum}
\end{align}
Substituting the explicit expressions for the components of $\vec v_k$, Eq.~\eqref{nastysum} requires evaluation of the sums
$2^{-n}\sum_k\cos^2{\varphi_k}$, $2^{-n}\sum_k\sin^2{\varphi_k}$, and $ 2^{-n}\sum_k\sin{\varphi_k}\cos{\varphi_k}$. We observe that for large $n$ and `typical' unitaries, where the phases $\varphi_k$ are uniformly distributed \cite{dat}, we can approximate those sums with integral averages of the trigonometric functions over the interval $\varphi\in[0,2\pi]$: $\langle \cos^2\rangle\!\simeq\!\langle\sin^2\rangle\!\simeq\!1/2,\langle\sin\cos\rangle\!\simeq\!0$. Then,
\begin{align}
W_{AB}\!\simeq\!\text{diag}\{v_0^2,v_0^2,v_0^2\!-\!v^2\}\!\Rightarrow\!\lambda_\text{max}(W_{AB})\!=\!v_0^2.
\end{align}
Finally, the conditions given above on $v_0,v$ can be used to express $v_0$ in terms of the qubit initial polarization, as $v_0^2=1/2(1+\sqrt{1-\mu^2})$. Substituting this in Eq.~\eqref{anale} yields the anticipated result of Eq.~\eqref{dqc1}.
As expected, the expression  increases monotonically with the ancilla polarization and is independent of the number of qubits in the register. This is in agreement with what predicted by using the quantum discord \cite{OZ,dat}.
\end{proof}
\end{Rem}
\ \\

\clearpage
 \end{widetext}


\begin{thebibliography}{99}
\bibitem{noteref}
Indeed, a nondegenerate qubit observable with fixed spectrum can be parametrized as $K_A=V_A(\alpha\sigma_{zA}+\beta\mathbb I_A)V_A^\dagger=\alpha\vec n\cdot \vec \sigma_A+\beta\mathbb I_A$, where $\alpha,\beta$ are constants ($\alpha\neq0$) while $\vec n$ varies on the unit sphere. Then $\mathcal I(\rho,K_A)=\alpha^2 \mathcal I(\rho,\vec n\cdot\vec \sigma_A)$, implying that all the ${\cal U}_A^\Lambda$'s are proportional to each other.
\end{thebibliography}
\end{document}